\documentclass[showpacs,twocolumn,aps,prb,amsmath,amssymb,superscriptaddress]{revtex4}
\usepackage{epstopdf}
\usepackage{graphicx}
\usepackage{dcolumn}
\usepackage{bm}
\usepackage{amsfonts}
\usepackage{amsmath}

\begin{document}
\renewcommand{\thefigure}{\arabic{figure}}
\def\be{\begin{equation}}
\def\ee{\end{equation}}
\def\ber{\begin{eqnarray}}
\def\eer{\end{eqnarray}}

\def\kv{{\bf k}}
\def\qv{{\bf q}}
\def\pv{{\bf p}}
\def\sigmav{{\bf \sigma}}
\def\tauv{{\bf \tau}}

\newcommand{\h}[1]{{\hat {#1}}}
\newcommand{\hdg}[1]{{\hat {#1}^\dagger}}
\newcommand{\bra}[1]{\left\langle{#1}\right|}
\newcommand{\ket}[1]{\left|{#1}\right\rangle}

\title{Electronic ground state properties of strained graphene }
\date{\today}
\author{H. Rostami}
\affiliation{School of Physics, Institute for Research in
Fundamental Sciences (IPM), Tehran 19395-5531, Iran}

\author{Reza Asgari}
\email{asgari@ipm.ir} \affiliation{School of Physics, Institute
for Research in Fundamental Sciences (IPM), Tehran 19395-5531,
Iran}

\begin{abstract}
We consider the effect of the Coulomb interaction in strained
graphene using tight-binding approximation together with the
Hartree-Fock interactions. The many-body energy dispersion relation, anisotropic Fermi velocity
renormalization and charge compressibility in the presence of 
uniaxial strain are calculated. We show that the quasiparticle
quantities are sensitive to homogenous strain and indeed, to its sign. The
charge compressibility is enhanced by stretching and
suppressed by compressing a graphene sheet. We find a reduction of
Fermi velocity renormalization along the direction of graphene deformation, in good agreement with
the recent experimental observation.

\end{abstract}

\pacs{71.10.Ca, 68.35.Gy, 71.18.+y} \maketitle

\section{Introduction}\label{sect:intro}
Graphene is a two-dimensional crystal of carbon atoms with
enormous interest on its unique features~\cite{novoselov}.
Graphene sheet has attracted considerable attention because of its
unusual electronic properties~\cite{kotov_arXiv_2010,
bostwick_science_2010} which follow from chiral band states and
because of potential applications~\cite{avouris}. The low energy
quasiparticle excitations energies in graphene are linearly
dispersing, described by Dirac cones at the edges of the first
Brillouin zone. Charge carriers move strictly in two-dimensions
since the graphene layer is just one atomic monolayer thick.

Graphene is amenable to external influences incorporating
mechanical deformation~\cite{ref:Seon, ref:Prieara}. Its band
structure does not change for realistic strains less than
$15\%$~\cite{ref:Prieara, lee_2008}. The influence of long-range
strains on the electronic properties is a unique feature of
graphene~\cite{ref:hopping, maes}. At low-energy spectrum, strains
give rise to a pseudomagnetic field which is added to the momentum
operators~\cite{vozmediano} and thus a gauge field couples to
electrons. The most evidence of the unusual way in which strains
affect the electronic states comes from scanning tunneling
microscope measurements of the electronic local density of states
of graphene grown on platinum~\cite{levy}. An average compression
of $10\%$ creates effective fields of the same order of magnitude with the value
observed in experiments~\cite{nima}.

Tuning the dynamics of massless carriers by appropriately designed
strain patterns~\cite{guinea} opens the way for novel applications
of graphene\cite{exp_strain}. Strain can be induced in graphene
either intentionally or naturally. Uniaxial strain can be
induced by bending the substrates on which graphene is elongated
without slippage. Elastic responses have been measured by pushing
a tip of atomic force microscopes on suspended graphene. The
presence of ripples in graphene samples and their influence on the
electronic properties are open problems in the field for instance
graphene on top of SiO$_2$ or SiC surface experiences a moderate
strain due to surface corrugations or lattice mismatch.

The Fermi velocity is an essential quantity in graphene because all the observable quantities depend on it. For the Dirac
electrons in pristine graphene, it was
shown~\cite{yafis,polini,asgari0,asgari,asgari_im, elias} that
interaction effects also become noticeable with decreasing density
in that the velocity is enhanced rather than suppressed, and that the
influence of interactions on the compressibility and the
spin-susceptibility changes sign. These qualitative differences
are due to {\it exchange interactions} between electrons near the
Fermi surface and electrons in the negative energy sea and to
interband contributions to Dirac electrons from charge and spin
fluctuations. Recent experiments~\cite{elias, eva} have been able
to measure the renormalized Fermi velocity and they claimed that
the Fermi velocity is no longer constant but increases by
decreasing the electron density. On the other hand, it was
shown~\cite{choi}, from first principles calculations, that the
group velocities under uniaxial strain exhibit a strong
anisotropy. As the uniaxial strain increases along a certain
direction, the Fermi velocity parallel to it decreases quickly and
vanishes eventually, whereas the Fermi velocity perpendicular to
it increases by as much as $25\%$. It was also shown that the work
function of strained graphene increases substantially as strain
increases.

The role of long-range electron-electron interactions in a
uniaxial strain on undoped graphene, has been recently
studied~\cite{sharma} using the renormalization group theory. They
showed that while for small interactions and anisotropy the system
restores the conventional isotropic Dirac liquid behavior, at
intermediate coupling both the anisotropy and interactions can
flow, in the renormalization group sense, toward large values thus
signaling the emergence of anisotropic excitonic states. On the
other hand, the dependence of the electron velocity on periodic
deformations of graphene has also been investigated by another
group~\cite{dugaev}. The authors showed that the Fermi velocity
anisotropy corresponds to the anisotropy of the quasiparticle
spectrum energy.

The positions of the Dirac points can be shifted in the opposite
directions by applying anisotropic strains. The reason for this is
that the time-reversal symmetry is preserved by strain in
graphene~\cite{sasaki}. However, the translational symmetry is
broken by lattice deformation.

Our aim in this work is to study the Coulomb interaction effects in
uniaxially strained graphene particulary the anisotropy of the Fermi
velocity renormalization and the charge compressibility within the
exchange interaction. In the Hartree-Fock theory, the positions of the Dirac points are
shifted in the opposite directions when the full Brillouin zone calculation is taken into account.
Our theory for strain dependence of
quasiparticle velocity renormalization in interacting Dirac
electron systems is motivated not only by fundamental many-body
considerations, but also by application and potential future
experiments in the strain engineering field.

The paper is organized as follows. In Sec. II we introduce
the formalism that will be used in calculating strained ground
state properties which includes the many-body effects by using
Hartree-Fock ( HF) approximation. In Sec. III we present our
analytical and numerical results for the renormalized Fermi
velocity and the charge compressibility in doped graphene sheets.
Sec. IV contains discussions and conclusions.

\section{Method and Theory}

\subsection{Noninteractacting strained graphene}

The electronic bands in graphene arise from the hybridization of
$p_z$ orbitals which are localized at each carbon atom. A good
description of the electron states is obtained by assuming that
electrons can hop between nearest-neighbor atoms. We consider a
two dimensional honeycomb lattice in which a $p_z$ electron hops
between lattice points and it is governed by the nearest-neighbor
tight-binding Hamiltonian

\begin{equation}\label{eq:HTB}
H_{0}=-\sum_{<ij>}{t_{ij} a^{\dagger}_i b_j}+H.c~.
\end{equation}
where $a_i$ and $b_j$ are fermion field operators in sublattices
$A$ and $B$, $t_{ij}$ are hopping integrals between lattice points
and will be different among different neighbors. The Fourier
component of the Hamiltonian is written as

\begin{equation}\label{eq:Hk}
H_{0}^k=\begin{pmatrix}\ 0\ \ \ f(\vec{k})\\ \\ f^*(\vec{k})\ \ \  0 \ \end{pmatrix}
\end{equation}

where the factor $f(\vec{k})$ is $-\sum_{i}{t_{i} \
e^{i\vec{k}\cdot\vec{\delta}^{(0)}_{i}}}$ with $t_i$ being the hopping
energy of two nearest neighbor atom. Notice that a lattice
deformation changes the off-diagonal terms in the Hamiltonian
which describe hopping between the two sublattices which make up
the honeycomb lattice. The induced scalar potential created by
strain is intensively suppressed by screening and thus we do not
take it into account.

\begin{figure}
\includegraphics[width=1\linewidth]{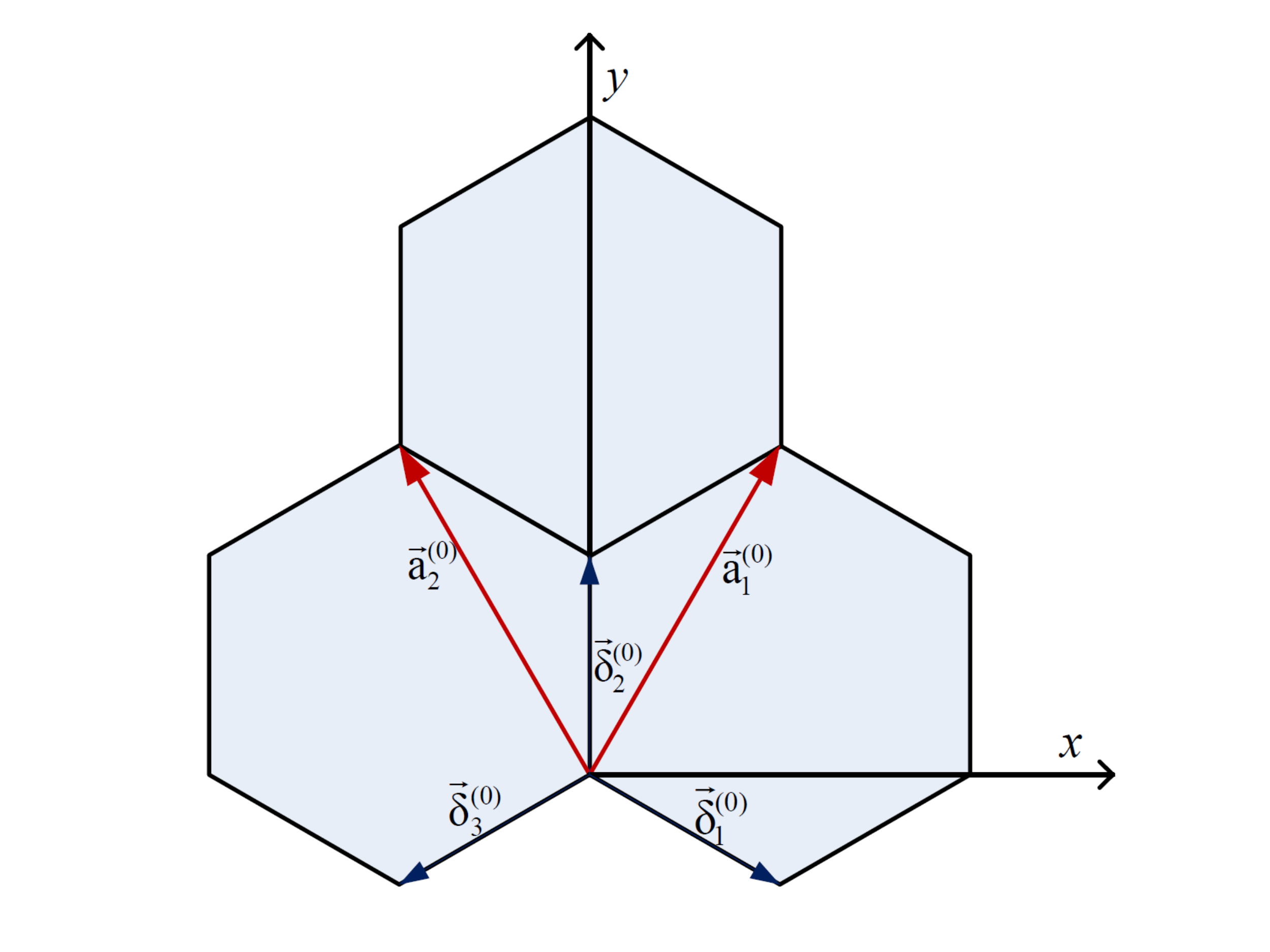}
\includegraphics[width=1\linewidth]{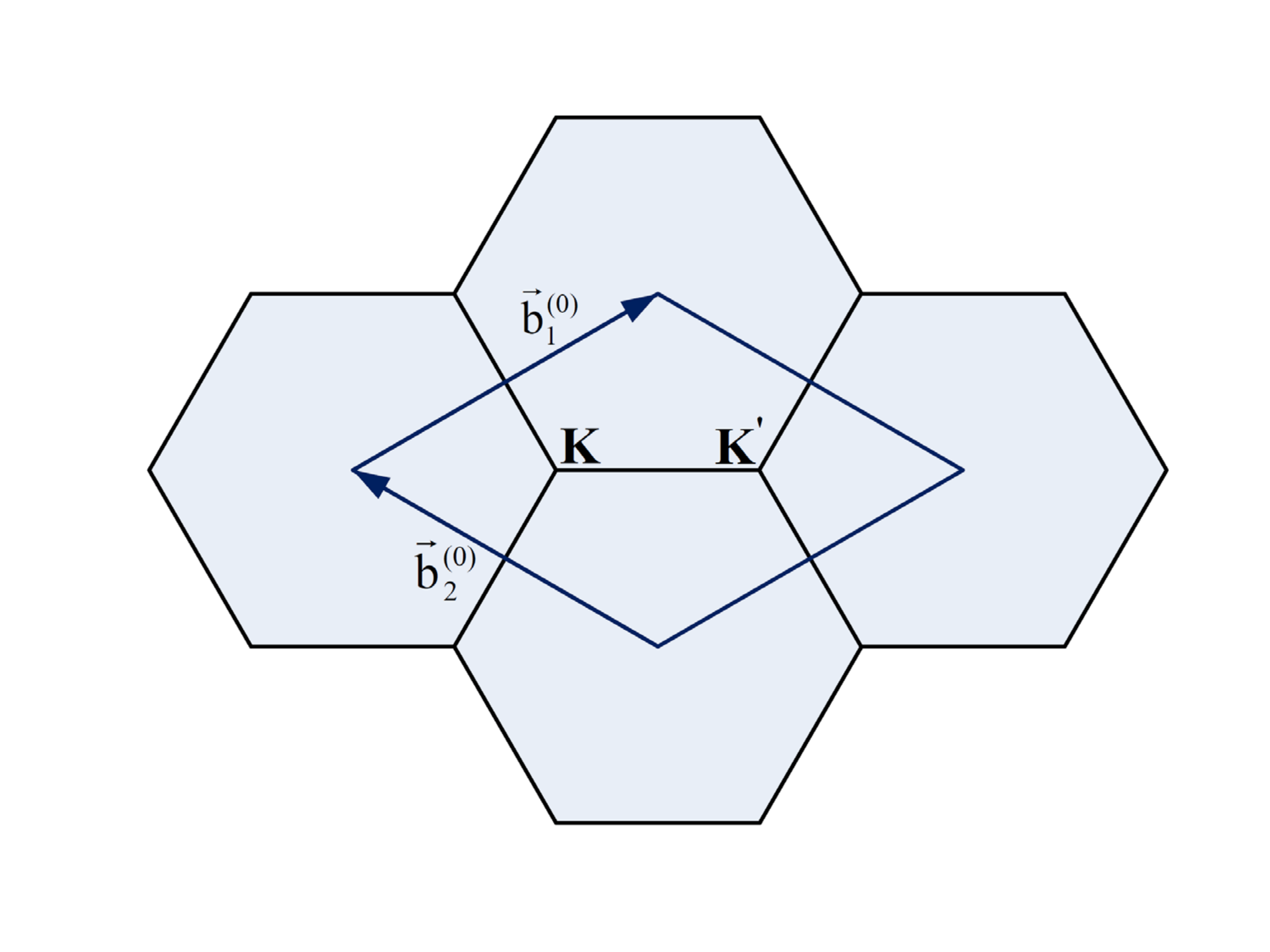}
\caption{(Color online) Top) Honeycomb lattice geometry incorporates three nearest-neighbor vectors and the
lattice primitive vectors of pristine graphene. Bottom) The first Brillouin zone of undeformed graphene incorporates two Dirac points with high symmetry. }
\label{fig:lattice}
\end{figure}

The honeycomb lattice geometry, three nearest-neighbor vectors and
the lattice primitive vectors of a pristine graphene are depicted in
Fig.~\ref{fig:lattice}. The lattice vectors are
${\vec{a}_1}^{(0)}=(1,\sqrt{3})\sqrt{3} a_0/2$ and
${\vec{a}_2}^{(0)}=(-1,\sqrt{3})\sqrt{3} a_0/2$ where
$a_0=1.42${\AA} is the $C-C$ equilibrium distance. The nearest
neighbor vectors are defined by

\begin{eqnarray}\label{eq:realspace}
{\vec{\delta}_1}^{(0)}&=&a_0(\frac{\sqrt{3}}{2},-\frac{1}{2})\nonumber\\
{\vec{\delta}_2}^{(0)}&=&a_0(0,1)\nonumber\\
{\vec{\delta}_3}^{(0)}&=&a_0(-\frac{\sqrt{3}}{2},-\frac{1}{2})
\end{eqnarray}

Reciprocal lattice basic vectors are also defined by
$\vec{a}^{(0)}_i\cdot\vec{b}^{(0)}_j=2\pi\delta_{ij}$. The Brillouin
zone which contains two Dirac points is shown in
Fig.~\ref{fig:lattice} (bottom) in which the position of $K-$point
in this perfect lattice is
$\vec{K}=\frac{4\pi}{3\sqrt{3}a_0}\hat{x}$. The nearest neighbor
vectors in strained lattice are defined~\cite{ref:hopping} by $
\vec{\delta^{'}}_{l,i}=\vec{\delta}^{(0)}_{l,i}+\sum_j{\epsilon_{ij}\vec{\delta}^{(0)}_{l,j}}$
and thus the modification of these distances distorts the
reciprocal lattice and it is easy to find the reciprocal vectors
under strain and the results are
\begin{eqnarray}
\vec{b^{'}}_1&=&\frac{2\pi}{3\sqrt{3} d}\left[3(1+\epsilon_{yy})-\sqrt{3}\epsilon_{xy},\sqrt{3}(1+\epsilon_{xx})-3\epsilon_{xy}\right]\nonumber\\
\vec{b^{'}}_2&=&\frac{2\pi}{3\sqrt{3} d}\left[-3(1+\epsilon_{yy})-\sqrt{3}\epsilon_{xy},\sqrt{3}(1+\epsilon_{xx})+3\epsilon_{xy}\right]~,\nonumber\\
\end{eqnarray}
where $d=1+\epsilon_{xx}+\epsilon_{yy}+O(\epsilon^2)$ and
$\epsilon_{yx}$ denotes a shear strain and being zero in the
uniaxial strain, respectively. The tensor for uniaxial strain along $x$
direction is
\begin{equation}
{\bf \epsilon}= \varepsilon \begin{pmatrix} 1&0\\0&-\nu \end{pmatrix}~,
\end{equation}
where $\nu=0.165$ is Poisson's ratio. Notice that we are
interested in uniform planar tension cases. Most importantly, the
Dirac point which is a symmetry point, will be shifted to a new
position. For homogeneous strain the position of Dirac point is
defined by the condition $f(K)=0$. After some straight
forward calculations, the new position of the Dirac
point is given by

\begin{eqnarray}
\vec{K^0_D}&=&\frac{1}{2\pi}(\theta_1 \vec{b}^{(0)}_1+\theta_2 \vec{b}^{(0)}_2)\label{eq:K}\\
\theta_1 &=&\cos^{-1}{(\frac{t^2_1-t^2_2-t^2_3}{2 t_2 t_3})}\label{eq:theta1}\nonumber\\
\theta_2 &=&-\cos^{-1}{(\frac{t^2_3-t^2_2-t^2_1}{2 t_1 t_2})}
\end{eqnarray}

Notice that the reciprocal vectors of unstrained graphene appear in Eq.~(\ref{eq:K}) instead of the distorted vectors,~\cite{ref:Yasumasa} and the reason of that originally comes from the tight-binding Hamiltonian and the definition of $f(k)$. In order to find the new position of the Dirac point, we only need to evaluate the hopping integrals. In fact, the
change of bond-length leads to different hopping integrals among
neighbors. Since the carbon atoms are out of equilibrium distance,
we set~\cite{ref:DAP}
\begin{equation}\label{eq:DAP}
t_{i}=t_0~e^{-\frac{9.1}{t_0}(\frac{|\vec{\delta'}_{i}|}{a_0}-1)}~,
\end{equation}
where $t_0=3.09eV$ gives rise to $v_F=10^6m/s$. This definition
satisfies a condition~\cite{neto} in which $\partial
t_{i}/\partial a=-6.4$ eV{\AA}$^{-1}$. The dispersion relation is
thus given by $E_0(\vec{k})=\pm |f(\vec{k})|$. For the sake of
completeness, we thus expand the form factor around the
$K^0_D$-point to obtain its low-energy expression and it results
in

\begin{equation}\label{eq:fkq}
f(\vec{K^0_D}+\vec{q})=-\sum_i t_{i} e^{i(\vec{K^0_D}+\vec{q})\cdot\vec{\delta}^{(0)}_{i}} \approx -i \sum_i {t_{i}\vec{q}\cdot\vec{\delta}^{(0)}_i~
 e^{i\vec{K^0_D}\cdot\vec{\delta}^{(0)}_{i}} }~.
\end{equation}
From now on, we drop $(0)$ index of the nearest-neighbors vectors
for simplicity. Equation~(\ref{eq:fkq}) is thus simplified as

\begin{figure}[ht]
\begin{center}
\includegraphics[width=1\linewidth]{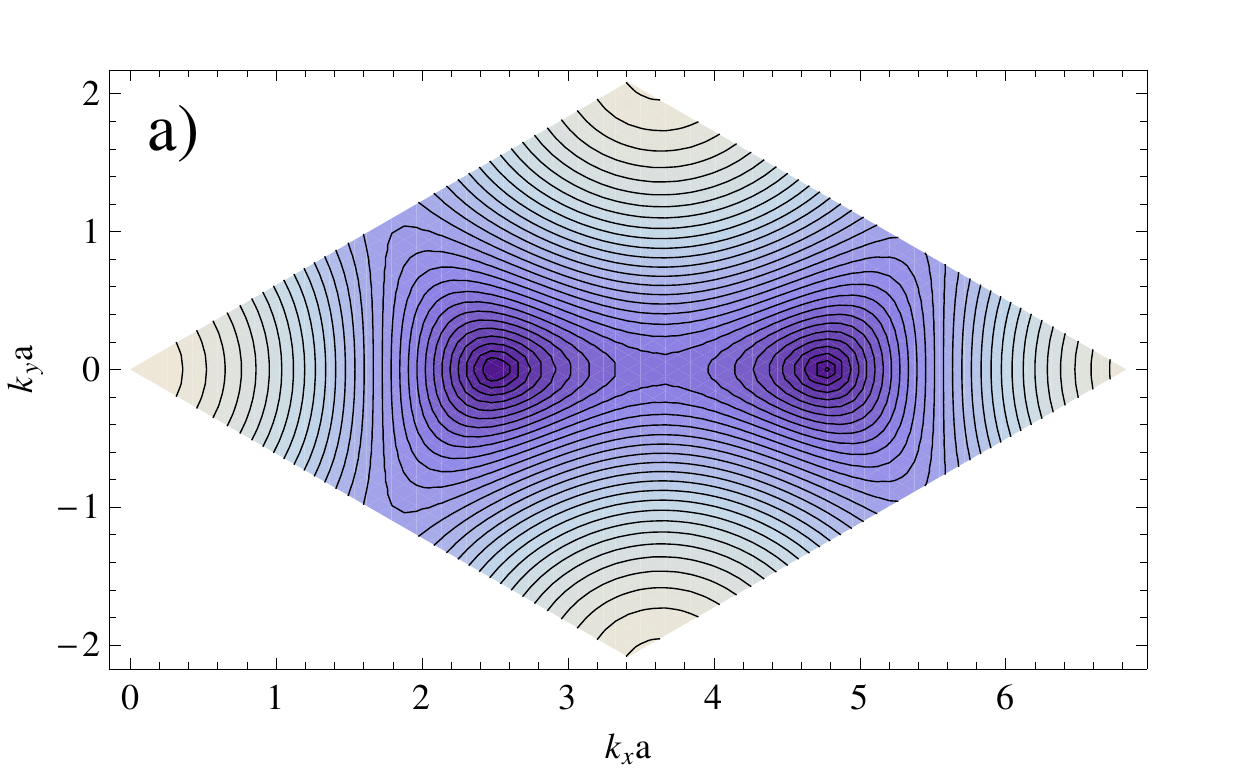}
\includegraphics[width=0.9\linewidth]{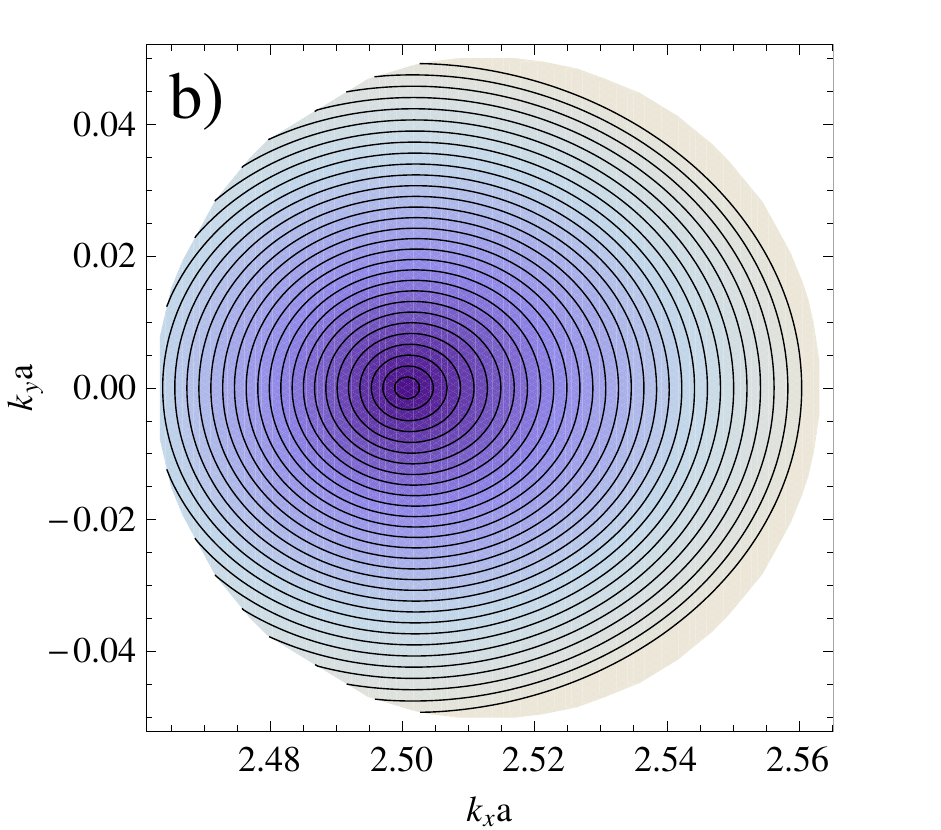}
\caption{ (Color online) a): Contour plot of the energy dispersion for strained and interacting graphene sheet and b)
a zoom of the energy dispersion around $K_D$-point. Notice that the center of box indicates the position of $K^0_D$ ( shift of the Dirac point in strained graphene
without the electron-electron interaction) and clearly $K_D$ moves to the left.
These results are calculated by considering $\varepsilon=0.05$, $\alpha_{ee}=0.5$ and $n=$10$^{12}$cm$^{-2}$.
The deviation from the linear dispersion relation associated to the trigonal warping effect
is clearly shown in (a) using the full expression of $f(k)$.}
\label{fig:ekfull3d}
\centering
\end{center}
\end{figure}

\begin{eqnarray}
|f(\vec{K^0_D}+\vec{q})|^2&\approx& t_1^2(\vec{q}.\vec{\delta}_1)^2+ t_2^2(\vec{q}.\vec{\delta}_2)^2+ t_3^2(\vec{q}.\vec{\delta}_3)^2\nonumber\\&+&
(\vec{q}.\vec{\delta}_1)(\vec{q}.\vec{\delta}_3)(t_3^2-t_1^2-t_2^2)\nonumber\\&+&
(\vec{q}.\vec{\delta}_1)(\vec{q}.\vec{\delta}_2)(t_2^2-t_1^2-t_3^2)\nonumber\\&+&
(\vec{q}.\vec{\delta}_2)(\vec{q}.\vec{\delta}_3)(t_1^2-t_2^2-t_3^2)
\end{eqnarray}
Therefore, the band-structure with arbitrary hopping integrals is
given by
\begin{equation}\label{e0}
E_0^2=|f(\vec{K_D}+\vec{q})|^2=\hbar^2(v_x^2q_x^2+v_y^2q_y^2)+c^2q_xq_y~,
\end{equation}
where $c^2=3\sqrt{3}a_0^2(t_3^2-t_1^2)/2$. Since we are interested
in uniaxially strained graphene along the zig-zag direction in which $t_1=t_3$, consequently
$c=0$. In this case, $v_x^2=3a_0^2(4t_1^2-t_2^2)/4\hbar^2$ and
$v_y^2=9a_0^2 t_2^2/4\hbar^2$. From this equation, it is obvious
that the cross section of the energy dispersion in strained
graphene is an elliptic shape. Accordingly, the electron velocity
is direction dependent and there is an anisotropy in Fermi
velocity~\cite{pellegrino}. Notice that the energy at $(q_{{\rm F}_x},0) $ and
$(0,q_{{\rm F}_y})$ is equal to the Fermi energy,
$\epsilon_{\rm F}$ with $k_{\rm F}=\sqrt{\pi n}$. Furthermore
the electron density reads as $n\pi=q_{{\rm F}_x} q_{{\rm F}_y}$
and we thus have
\begin{equation}
\epsilon_{\rm F}=\hbar\sqrt{v_x v_y}\sqrt{n\pi}
\end{equation}
where $q_{{\rm F}_x}=k_{\rm F}\delta$, $q_{{\rm F}_y}=k_{\rm
F}\delta^{-1}$ and $\delta=\sqrt{v_y/v_x}$.

\begin{figure}
\begin{center}
\includegraphics[width=1\linewidth]{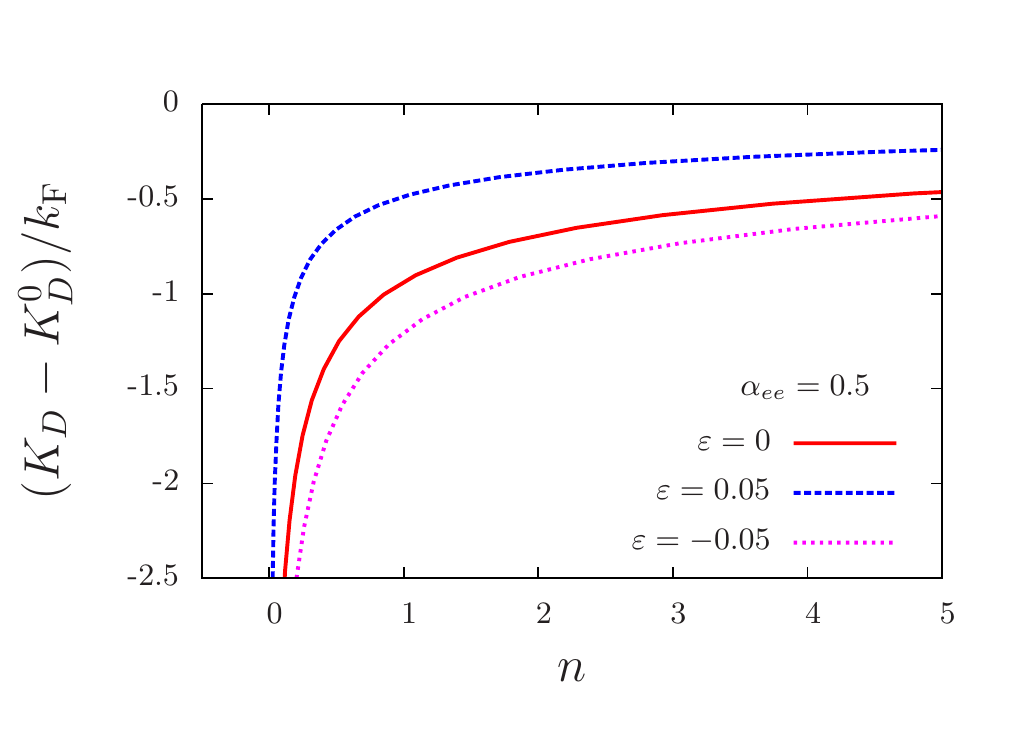}
\caption{(Color online) Dirac point shift in respect to the noninteracting case as a function of the electron density ( in units of 10$^{12}$ cm$^{-2}$) for strained and interacting graphene sheet.
The Dirac point moves noticeably at the low electron density and tends to zero linearly in the high electron density. }
\label{fig:KD}
\end{center}
\end{figure}

Note that the electron density will change when we apply strain to a
honeycomb lattice because the primitive cell area changes due to
the strain as
$\Omega^\prime=\Omega(1+\epsilon_{xx}+\epsilon_{yy})=\Omega(1+\varepsilon(1-\nu))$
so that

\begin{equation}
n=\frac{n_0}{1+\varepsilon(1-\nu)}
\end{equation}
where $n_0$ is the electron density in unstrained graphene.

\subsection{Interacting electrons in Hartree-Fock approximation }

We consider the interaction of quasiparticles by using the leading
diagram approximation which is the exchange interaction. In this
sense, when interactions are treated in a mean-field approximation
the Hamiltonian is written as ${\cal H}=H_0+H_{MF}$
where~\cite{ref:Borghi}

\begin{equation}\label{eq:hamil}
H_{MF}=-\frac{2}{S}\sum_{k k^\prime}\sum_{\alpha \beta}{v_{\vec{k}-\vec{k^\prime}}
 \rho_{\alpha\beta}(\vec{k^\prime}) \psi^\dagger_{k\alpha}\psi_{k\beta}}~,
\end{equation}
factor $2$ stands for spin degeneracy and $\rho$ is a density
operator given by $\rho_{\alpha\beta}(\vec{k})=
(n^{(0)}_{\vec{k}+}|\psi_0^+\rangle\langle\psi_0^+|+n^{(0)}_{\vec{k}-}|\psi_0^-\rangle\langle\psi_0^-|)_{\alpha\beta}
$ in which $n^{(0)}_{\vec{k}\pm}$ are noninteracting band
occupation factors, $v_q$ is the Fourier transformation of the
interparticle interaction. $|\psi_0^\pm\rangle$, the eigenvectors
of noninteracting Hamiltonian when the full $f(\vec{k})$
expression is considered, are given by

\begin{equation}\label{eq:psi}
|\psi_0^\pm\rangle =\frac{1}{\sqrt{2}}\begin{pmatrix}1 \\ \pm e^{-i\phi_f}\end{pmatrix}
\end{equation}
where $\exp[{i\phi_f}(\vec{k})]=f(\vec{k})/|f(\vec{k})|$. After
substituting Eq.~\ref{eq:psi} into the pseudospin density matrix,
it can be decomposed into the charge and pseudospin-density
contributions
\begin{equation}
\rho_{\alpha\beta}(\vec{k})=f_+(\vec{k})\delta_{\alpha\beta}+f_-(\vec{k}) \hat{n}_f.\vec{\sigma}_{\alpha\beta}
\end{equation}

where the unit vector is
$\hat{n}_f=\hat{x}\cos(\phi_f)-\hat{y}\sin(\phi_f)$ and we have
introduced a short-hand notation $f_{\pm}=(n^{(0)}_{k+}\pm
n^{(0)}_{k-})/2$.

It is easy to see that the Hamiltonian consists of a
momentum-dependent pseudospin effective magnetic field which acts
in the direction of momentum $k$. The band eigenstates in the
positive and negative energy bands have their pseudospins either
aligned or opposed to the direction of momentum. Therefore, the
interaction mean-field Hamiltonian changes to
\begin{widetext}
\begin{eqnarray}\label{eq:Ham}
H_{HF}&=&\sum_{k\alpha\beta}{\psi^\dagger_{k\alpha}\{\ \delta_{\alpha\beta} B_0(\vec{k})+\vec{\sigma}_{\alpha\beta}.\vec{B}(\vec{k})\}\psi_{k\beta}}\nonumber\\
B_0(\vec{k})&=&-2\int \frac{d \vec{k'}}{(2\pi)^2} v_{\vec{k}-\vec{k^\prime}} f_+(\vec{k^\prime})\nonumber\\
\vec{B}(\vec{k})&=&\hat{x}\Re e f(\vec{k})-\hat{y}\Im m f(\vec{k})-2\int \frac{d\vec{k'}}{(2\pi)^2} v_{\vec{k}-\vec{k'}}(\hat{x}\cos(\phi_f(\vec{k'}))-\hat{y}\sin(\phi_f(\vec{k'}))) f_-(\vec{k'})
\end{eqnarray}
\end{widetext}
where the integral term of $\vec{B}(\vec{k})$ is the exchange
field and the sum of the first two terms are the band-structure
pseudospin magnetic field. It should be noticed that Eq.~(\ref{eq:Ham}) is our main theoretical result. At zero temperature and for an n-doped
graphene layer,
$f_+(\vec{k})=\frac{1}{2}\theta(\epsilon_F-\epsilon_{\vec{k}})+\frac{1}{2}$
and $
f_-(\vec{k})=-\frac{1}{2}\theta(\epsilon_{\vec{k}}-\epsilon_F)$
where $\epsilon_F$ and $\epsilon_{\vec{k}}$ are the Fermi energy
and the noninteracting energy dispersion, respectively. To account
partially for screening and to avoid the well-known Fermi velocity
artifact of mean-field theory in systems with long-range
interactions, we have used an interaction potential including
Thomas-Fermi screening

\begin{equation}
v_{k-k^\prime}=\frac{2\pi e^2}{\overline{\epsilon} (q_{_{TF}}+|\vec{k}-\vec{k^\prime}|)}~,
\end{equation}
where $q_{_{TF}}=\alpha_{ee} k_F $ is the Thomas-Fermi screening
vector and $\alpha_{ee}=e^2/({\overline{\epsilon}} \hbar v_F)$ is the graphene
coupling constant, which is density independent. ${\overline{\epsilon}}$ is an
average dielectric constant of the surrounding medium.

Notice that in our calculations, all integrals in the phase space
are taken over the first Brillouin zone that is, a primitive unit
cell of the reciprocal lattice ( see Fig.~\ref{fig:lattice}
(bottom). For the sake of simplicity, it is convenient to work in a
frame $(k_1,k_2)$ where $k_1$ and $k_2$ are along $\vec{b'}_1$ and
$-\vec{b'}_2$, respectively. Therefore, it is accomplished by
changing $k_x=b^{'}_{1,x}(k_1+k_2)/|\vec{b'}_1|$ and
$k_y=b^{'}_{1,y}(k_1-k_2)/|\vec{b'}_1|$ where the Jacobian is
\begin{eqnarray}
J=2\frac{b^{'}_{1,x}b^{'}_{1,y}}{|\vec{b'}_1|^2 }=6\sqrt{3}\frac{(1-\nu\varepsilon)(1+\varepsilon)}{9(1-\nu\varepsilon)^2+3(1+\varepsilon)^2}
\end{eqnarray}

The real and imaginary part of $f(k)$ are given by $\Re e
f(\vec{k})=-\sum_l t_l \cos(\vec{k}\cdot\vec{\delta}_l)$ and $ \Im
m f(\vec{k})=-\sum_lt_l \sin(\vec{k}\cdot\vec{\delta}_l)$,
respectively. After having all ingredients, we are able to write
the energy dispersion in HF approximation
\begin{equation}\label{eq:ek}
E_\pm(\vec{k})=B_0(k)\pm \sqrt{B_x(\vec{k})^2+B_y(\vec{k})^2}~.
\end{equation}

Once $B_0(k)$ and $\vec{B}(\vec{k})$ are obtained, the ground
state quantities can be calculated. Due to the exchange
interaction, the Dirac points are shifted and a new Dirac point,
$K_D$ is found by searching the zeroth of Eq.~(\ref{eq:ek}).

\begin{figure}
\includegraphics[width=1\linewidth]{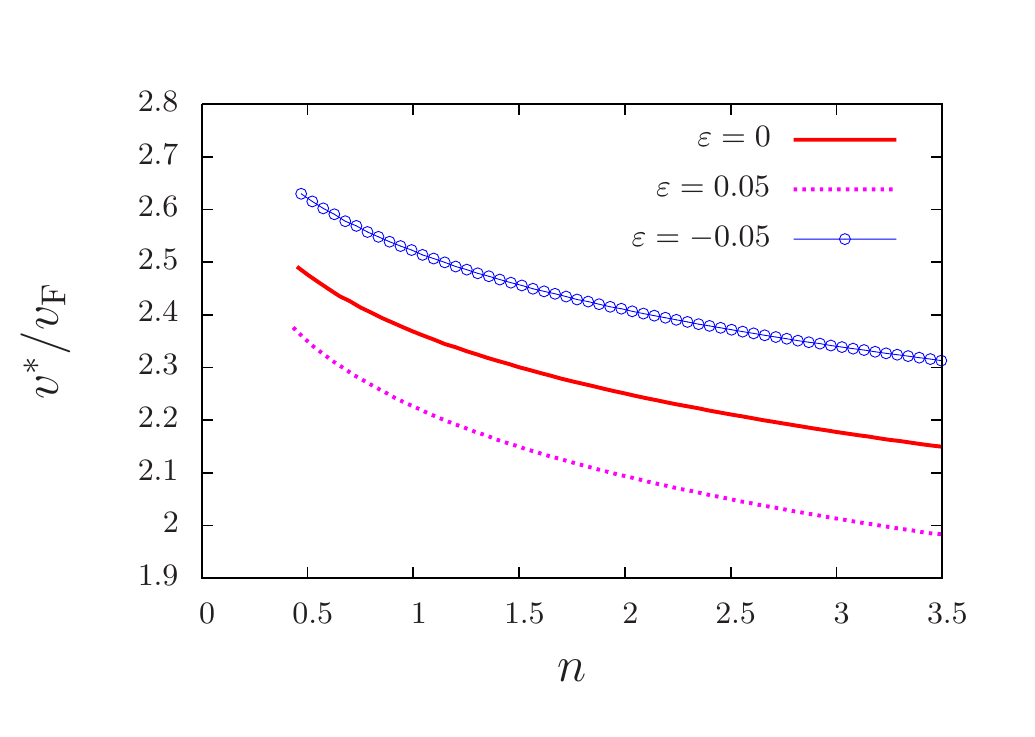}
\caption{(Color online) Strain dependence of the renormalized velocity at $(q_{{\rm F}_x},0)$ point, scaled by that of a noninteracting velocity
as a function of the electron density ( in units of 10$^{12}$ cm$^{-2}$) for different strain values and $\alpha_{ee}=0.9$.
Since the dispersion relation has an elliptic shape,
the Fermi velocity has also an anisotropic form.
For the positive strain values, stretching the
sample, the renormalized Fermi velocity at $(q_{{\rm F}_x},0)$ point
suppresses however, the renormalized Fermi velocity enhances by
compressing the sheet.
Importantly, the strain dependence of the Fermi velocity behaves differently as the sign of the strain is changed.}
\label{fig:vxpvf}
\end{figure}

\subsection{Anisotropy renormalized Fermi velocity and charge compressibility}

The effective Fermi velocity is an important concept in Landau's
Fermi liquid theory since it provides a direct measure of the
many-body interactions in the electron system. The low-energy
electronic excitations in graphene are described by a massless
Dirac Hamiltonian that is able to explain many transport
properties. The Fermi velocity is the only parameter that appears in
the model Hamiltonian and it plays the same role as the effective
mass~\cite{asgari2} in the standard Landau's Fermi liquid theory.
The renormalized Fermi velocity can be calculated when the ground
state energy dispersion as a function of $\vec{k}$ is obtained.
The renormalized Fermi velocity can be
expressed~\cite{Giuliani_and_Vignale} in terms of the wavevector
derivative of the charge carrier dispersion relation evaluated at
the Fermi surface

\begin{equation}
\vec{v^*}=\frac{1}{\hbar} \vec{\nabla} E(\vec{k}) |_{\vec{k}=\vec{K}_D+\vec{q}_F}~,
\end{equation}
and therefore, the Fermi velocity component along $i$th-direction
is given by
\begin{equation}
\hbar v_i^*=\frac{\partial B_0(k)}{\partial k_i}+\frac{B_x(k)\frac{\partial B_x(k)}{\partial k_i}+B_y(k)\frac{\partial B_y(k)}
{\partial k_i}}{\sqrt{B_x(k)^2+B_y(k)^2}}|_{\vec{k}=\vec{K}_D+\vec{q}_F}\\
\end{equation}

Once the ground state is obtained the compressibility, $\kappa$ can
be easily calculated from

\begin{equation}
\frac{1}{n^2 \kappa}=\frac{\partial \mu}{\partial n}
\end{equation}
where $\mu$ is the chemical potential of quasiparticles and incorporates
the kinetic energy and the first order exchange interaction in the
presence of the uniaxial strain. Therefore, it is easy to get the
following expression.
\begin{equation}
\frac{\partial \mu}{\partial n}=\frac{\partial k_{\rm F}}{\partial n}\frac{\partial \mu}{\partial k_{\rm F}}~,
\end{equation}
in which $\partial k_{\rm F}/\partial n=k_{\rm F}/(2 n)$. Notice
that the compressibility of the noninteracting system at low-
energy is given by $\kappa_0=2/(n \hbar
\sqrt{v_xv_y}\sqrt{\pi n} )$. The expression reduces to the
well-known expression of the noninteracting compressibility by
setting $v_x=v_y$.

\section{Numerical Results}

In this section, we present our calculations for the ground-state
properties of graphene in the presence of uniaxial strain which we
model as mentioned above. The shift of Dirac points,
the anisotropy of the Fermi velocity renormalization and inverse
compressibility $1/(n^2 \kappa)$ are calculated by using the
theoretical models and we compared the latter with the
recent experimental measurements. In all numerical calculations
presented here, we take into account the full Brillouin zone by
using the expression of $f(\vec{k})$. In this case, accurate
calculations require dense $k-$ point sampling near the Dirac
point and thus we use a dense adaptive sampling model the same as
used in Ref.~[\onlinecite{jung}]. We have also dropped the
constant $1/2$ contribution from the integrand of $B_0(\vec{k})$
and the reason is that the latter term produces a constant energy
contribution.

\begin{figure}
\includegraphics[width=1\linewidth]{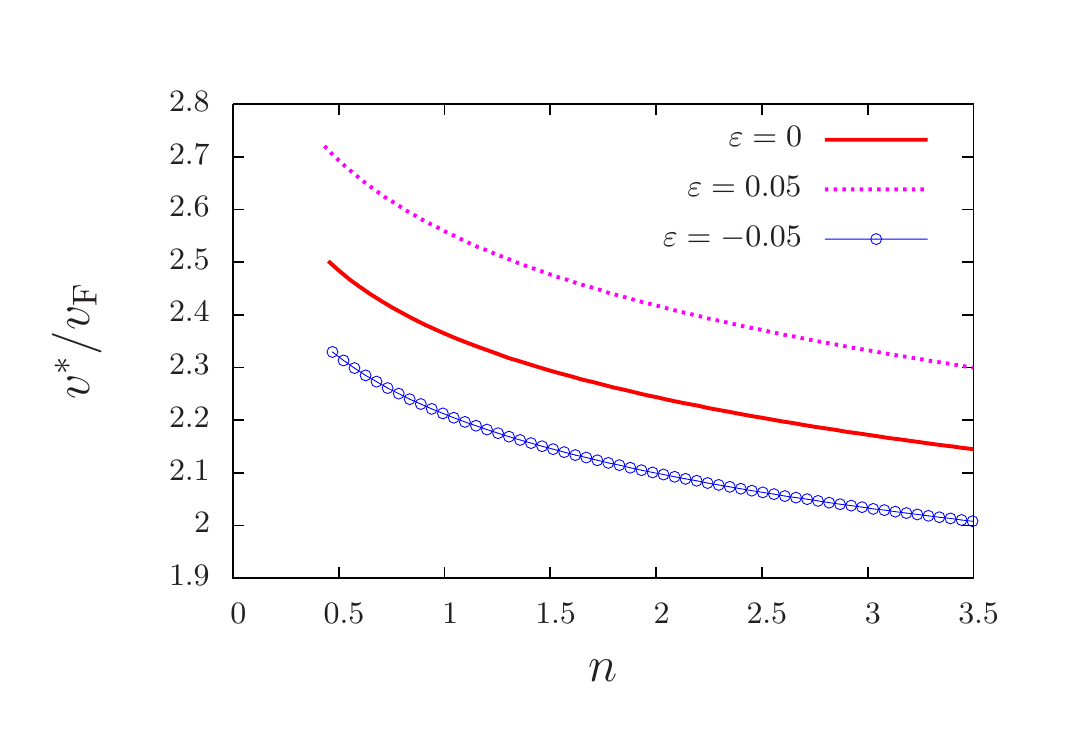}
\caption{(Color online) Strain dependence of the renormalized velocity at $(0,q_{{\rm F}_y})$ point, scaled by that of a noninteracting velocity
as a function of the electron density ( in units of 10$^{12}$ cm$^{-2}$) for different strain values and $\alpha_{ee}=0.9$.
For the positive strain values, stretching the
sample, the renormalized Fermi velocity at $(0,q_{{\rm F}_y})$ point
is enhanced however, the renormalized Fermi velocity is suppressed by
compressing the sheet. The strain dependence of the Fermi velocity behaves differently with
those results obtained at $(q_{{\rm F}_x},0)$.}
\label{fig:vxpvf}
\end{figure}

\begin{figure}
\includegraphics[width=1\linewidth]{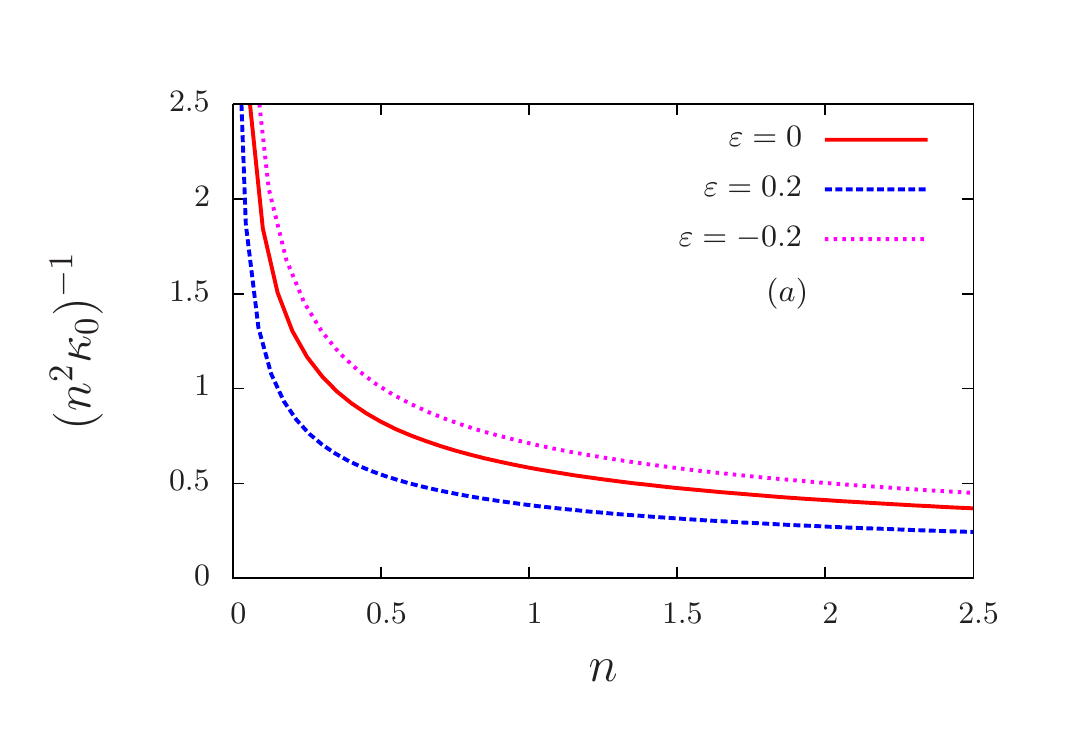}
\includegraphics[width=1\linewidth]{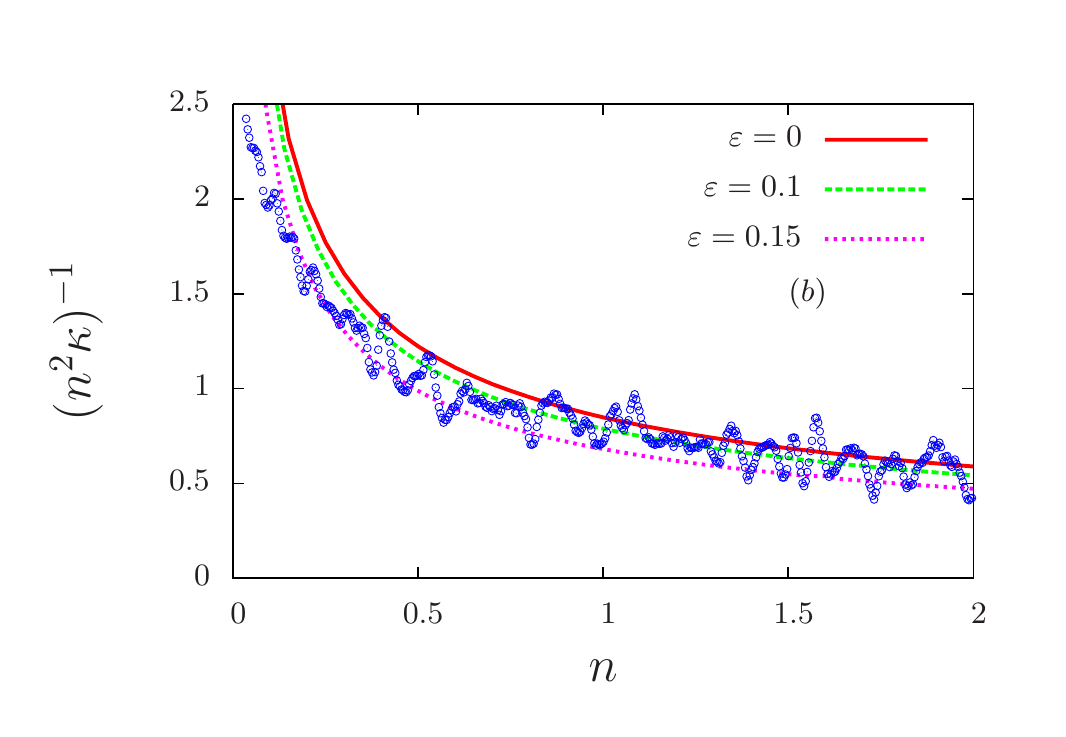}
\caption{(Color online) The inverse compressibility
$[n^2\kappa]^{-1}=\partial \mu/\partial n$ (in units of ${\rm
meV}~10^{-10}{\rm cm}^{2}$) as a function of the electron density
(in units of $10^{12}~{\rm cm}^{-2}$) for different strain values for (a) a noninteracting and (b) an interacting massless
Dirac electron system in graphene sheets. The symbols are the
experimental data by Martin {\it et al.}~\cite{martin}.
The strain dependence of the charge compressibility behaves differently as the sign of the uniaxial strain is changed.
The coupling constant is $\alpha_{ee}=0.26$. }
\label{fig:comp}
\end{figure}

As we have discussed in the previous section, the Dirac point
moves to a new position by taking into account the full $f(k)$ in
strained and interacting graphene. The contour plot of the
massless Dirac excitation energy is shown in Fig.~2 for
$\alpha_{ee}=0.5$. Notice that the center of the Fig.~2(b)
indicates the position of $K^0_D$ ( the Dirac shift of strained
graphene without the electron-electron interaction) and clearly
$K_D$ moves to the left. It is also clear from the figure that the
Dirac points, defined by $E_+(\vec{K}_D)=E_-(\vec{K}_D)$, are not
exactly on the $K^{0}_D$ and $K^{'0}_D$ points. It is easy to show that
$B_y$ vanishes along $x$ direction for the uniaxial strain. Since
$B_0(K_D)$ does not vanish, we might define an energy shift so
that to get vanishing dispersion relation at the Dirac point.
Therefore, the excitation energy is redefined as
$\mathcal{E}(\vec{k})=E(\vec{k})-E(\vec{K_D})$. A deviation from the
linear dispersion relation associated to the trigonal warping
effect is clearly shown in Fig.~2(a) using the full expression
of $f(\vec{k})$.

It is noticed that in a noninteracting graphene system, the Dirac
point is given by Eq.~(\ref{eq:K}) and it differs with
$K=4\pi/(3\sqrt{3} a_0)$ when the lattice is unperturbed. Due to
the charge-charge interaction, the Dirac point changes and the
shift is found by searching the zeroth of Eq.~(\ref{eq:ek}). The
Dirac shift, $K_D-K^0_D$ scaled by $k_{\rm F}$, as a function of
the electron density ( in units of 10$^{12}$ cm$^{-2}$) is
depicted in Fig.~\ref{fig:KD}. The shift is noticeable at the low
electron density region and tends to zero linearly in the high
densities. It is worthwhile mentioning that the Dirac shift value
is enhanced when a graphene is compressed however, it is suppressed when a graphene sheet is stretched. In addition, the shift
disappears~\cite{ref:Borghi} in the linear limit of $f(\vec{k})$
expression.

Our results for the strain dependence of the Dirac electron velocity,
${v^\star}/v_{\rm F}$, at $(q_{{\rm F}_x},0)$ point as a function
of the electron density $n$ are demonstrated in Fig.~4. Since the
dispersion relation has an elliptic shape, the Fermi velocity has an
anisotropic form. The numerical results are shown for
different strain values, $\varepsilon$ and for the graphene
coupling constant, $\alpha_{ee}=0.9$. For the positive strain
values, by stretching the sample, the renormalized Fermi velocity at
$(q_{{\rm F}_x},0)$ point suppresses, however the renormalized
Fermi velocity enhances by compressing the sheet. These behaviors
are based on the effective hopping terms in the presence of the
strain. Besides, the Fermi velocity decreases by increasing the electron density due to the exchange interactions between electrons near the Fermi surface. The strain dependence of the Dirac electron
velocity at $(0,q_{{\rm F}_y})$ point, on the other hand, behaves differently with
those results obtained at $(q_{{\rm F}_x},0)$ point and our
numerical results are shown in Fig.~5. As the uniaxial strain
increases along the dirction of graphene deformation, the Fermi velocity in this direction decreases
quickly, whereas the Fermi velocity perpendicular to it increases.

Recently, the uniaxial deformation of graphene on the
unidirectionally modulated SiC steps shaped substrate is studied by using
angular resolved photoemission spectroscopy~\cite{kan}. A
reduction of the Fermi velocity along the direction of graphene deformation, $v_x$ was
observed, while in another direction the changes were not
significant. Our numerical results on the Fermi velocity along
the direction of graphene deformation are in good agreement with the recent
observations. The reason for the insensitivity of the
Fermi velocity perpendicular to the graphene deformation, $v_y$ is probably due to
the fact that the sample is shaped as a supperlattice along the direction of deformation. The sample can be roughly realized as a combination of a pristine supperlattice graphene together with a uniaxial strained ribbon. For the pristine supperlattice, $v_y$ decreases and $v_x$ remains unchanged. On the other hand, we show that $v_y$ increases when graphene is stretched and therefore there should be a cancellation between two contributions and it leads to insensitivity of the
Fermi velocity perpendicular to the direction of graphene deformation in agreement with the experimental measurements.

The local compressibility of graphene, on the other hand, has been
measured ~\cite{martin} using a scannable single electron
transistor. From the theoretical point of view, the
compressibility was also studied by different groups at zero
temperature~\cite{peres, yafis,asgari_im} and also at finite
temperature~\cite{ramezanali}. In order to calculate the charge
compressibility in the presence of strain, we do need to calculate
the derivative of the total energy as a function of the electron
density. It should be noted that $\partial B_0(\vec{k})/\partial
k_{\rm F}$ and $\partial {\vec B}(\vec{k})/\partial k_{\rm F}$
change slightly by changing $\phi=\tan^{-1}(k_y/k_x)$ which is a consequence of the
elliptic shape of the Fermi surface and the $\phi$ dependence
tends to zero for small $\varepsilon$ values. Our numerical
compressibility is obtained by taking an average over $\phi$ and
the results are shown in Fig.~\ref{fig:comp}. The strain
dependence of the compressibility for a noninteracting system is
also demonstrated in Fig.~\ref{fig:comp}(a). For the
noninteracting case, it is easy to find that $\sqrt{v_x
v_y}\propto v_{\rm F}(1-9.1(1-\nu)\varepsilon/(8t_0))$ and thus for
$\varepsilon>0$ it gives rise to a reduction of the inverse of the
compressibility. Importantly, the charge compressibility decreases
(increases) by compressing ( stretching) a graphene sheet. Also the
asymmetric response to the positive and negative strain is evident
in the charge compressibility as is evident in the renormalized Fermi
velocity.

In Fig.~\ref{fig:comp}(b) we compare our theoretical predictions
for the inverse compressibility of doped graphene with the
experimental results of Martin {\it et al.}~\cite{martin} as a
function of the electron density in units of $10^{12}~{\rm
cm}^{-2}$. Martin {\it et al.}~\cite{martin} fitted the
experimental inverse compressibility, $(n^2 \kappa)^{-1}$ to the
kinetic term using a single parameter Fermi velocity which is
larger than the bare Fermi velocity. Note that the kinetic term in
graphene has the same density dependence as the leading exchange
and correlations terms~\cite{asgari_im}.

As it is clear in Fig.~\ref{fig:comp}(b) the inverse
compressibility of an interacting system is higher than the experimental
value. By increasing the uniaxial strain effects, i.e., increasing
the constant strength, $\varepsilon$ our theoretical results move
down. Therefore, including the exchange and strain effects in our
theory, gives results very close to experimental data.

\section{Conclusion}\label{sect:concl}

We have studied the ground state properties of a graphene sheet
within the Hartree-Fock theory incorporating the uniaxial strain
in the system. We have shown that the quasiparticle properties could be strongly strain dependent and substantially different than the usual pristine
graphene sheet. The Dirac points move due to the strain and
the anisotropy of Fermi velocity renormalization is obtained. The Dirac
electron Fermi velocity is highly dependent on the sign of strain
( stretching or compressing) even if the strain is negligibly
small. The renormalized electron velocity along the direction of graphene
deformation decreases with stretching however, it increases by
compressing the graphene sample. The Fermi velocity along a
direction perpendicular to graphene deformation behaves in a different way. We find the reduction of Fermi velocity
along the direction of graphene deformation to be in good agreement with the recent
experimental observation.

Our calculations of inverse compressibility compared with recent
experimental results of Martin {\it et al}.\cite{martin}
demonstrate the important influence of strain on the thermodynamic quantities in a  graphene sample.

We remark that in a very small density region, the system is
highly inhomogeneous and the effect of strain might be very
essential. A model going beyond the Hartree-Fock approximation is necessary to
account for increasing correlation effects at low density.

\acknowledgments

We thank S. Abedinpour for useful discussions. This work was
supported by IPM.

\end{document}